\documentstyle[aps,preprint,tighten]{revtex}
\def\beq{\begin{equation}}
\def\eeq{\end{equation}}
\def\beqa{\begin{eqnarray}}
\def\eeqa{\end{eqnarray}}

\def\GeV{\nobreak\,\mbox{GeV}}

\begin{document}
\title{\sc  $g_{ND\Lambda_{c}}$ from QCD Sum Rules}
\author {F.S. Navarra\thanks{e-mail: navarra@if.usp.br} and
\ M. Nielsen\thanks{e-mail: mnielsen@if.usp.br}\\
{\it Instituto de F\'{\i}sica, Universidade de S\~{a}o Paulo}\\
{\it C.P. 66318,  05389-970 S\~{a}o Paulo, SP, Brazil}}
\maketitle
\vspace{1cm}

\begin{abstract}
The $ND\Lambda_c$ coupling constant is evaluated in a full
QCD sum rule calculation. We study the Borel sum rule for the three
point function of one pseudoscalar one nucleon and one $\Lambda_c$ current
up to order seven in the operator product expansion. The Borel transform
is performed with respect to the nucleon and $\Lambda_c$ momenta, which 
are taken to be equal, whereas the momentum $q^2$ of the pseudoscalar
vertex is taken to be zero. This coupling constant is relevant in the
meson cloud description of the nucleon which has been recently used 
to explain exotic events observed by the 
H1 and ZEUS Collaborations at HERA.
\\
PACS numbers 14.20.Lq~~12.38.Lg~~14.65.Dw
\\

\end{abstract}

\vspace{1cm}

Many explanations have been  proposed for 
the recent observation by H1 and ZEUS Collaborations at HERA of an excess
of events at large  $x$ and $Q^2$ in deep inelastic scattering (DIS).  
Whereas most
explanations have concentrated on physics beyond the standard model, some
of them were based on more conventional mechanisms. One of these was 
suggested by Melnitchouk and Thomas \cite{mt}. They pointed out that an
enhanced charm quark distribution in the nucleon sea
 at large $x$ could account for the 
observed effect. This hard component of the $c$ distribution would have to
be generated non-perturbatively, since the usual charm quark distribution 
arising from gluon bremsstrahlung alone looks like a typical soft sea 
distribution, namely peaks at $x\rightarrow0$ and is negligible beyond
$x\simeq0.4$. In the works \cite{nnnt,pnndb} it was suggested that
the
charmed sea may arise from the quantum fluctuation of the nucleon
to a virtual $D^-\,\Lambda_c$ configuration. A natural prediction of this
picture are non-symmetric $c$ and $\overline c$ distributions, the latter
being much harder. Following this approach Melnitchouk and Thomas 
studied the effects of this charmed meson cloud on the DIS cross
sections measured at HERA.

In the framework of the standard meson cloud model the anti-charm quark
distribution in the nucleon, $\overline c_N$, is given by the
convolution of the valence $\overline c$ momentum distribution in
the ${D^-}$ meson, ${\overline c}_D (x)$, with the 
momentum distribution of this meson in the nucleon $ f_{D} (y)$ :
\begin{equation}
 {\overline c}_N (x) = \int_{x}^{1}
dy\, f_{D} (y)\, \frac{1}{y} \, {\overline c}_D (\frac{x}{y})\; .  
\label{cn}
\end{equation}                       
The virtual ${D^-}$ meson distribution in the nucleon cloud
which characterizes its probability of carrying a fraction $y$ of the
nucleon momentum in the infinite momentum frame is given by \cite{mt}
\begin{equation}
f_{D} (y) = \frac{g^2_{ND\Lambda_c}}{16 \pi^2}\, 
\int_{0}^{\infty}dk^2_{\bot} \, 
\frac{F^2 (k^2_{\bot},y)}{y\,(1-y)\,(s_{D\Lambda_c}-M_N^2)^2}\, \left(
\frac{k^2_{\bot}+[M_{\Lambda_c}-(1-y)M_N]^2}{1-y}\right)
\label{f}
\end{equation}
where $s_{D\Lambda_c}= 
(k^2_{\bot}+m^2_D)/y+(k^2_{\bot}+M_{\Lambda_c}^2)/(1-y)$.  
In the above equation $F(k^2_{\bot},y)$ is a
form factor at the $DN\Lambda_c$ vertex \cite{mt}.

In the heavy quark effective theory \cite{man} it is assumed that the
heavy quark interacting with light constituents, inside a hadronic
bound state, exchanges momenta much smaller than its mass. Therefore,
to a good approximation, the heavy quark moves with the velocity
of the charmed hadron. Being almost on-shell, the heavy quark carries
almost the whole momentum of the hadron. These considerations suggest
that the $\overline c$ distribution in the
$D^-$ meson is expected to be quite hard and may be approximated 
by a delta function:
\begin{eqnarray}
{\overline c}_D (\frac{x}{y}) &\simeq& \delta\,(\frac{x}{y}-1) 
\,= \, x\, \delta\,(x-y)\; .
\label{del}
\end{eqnarray}

From eqs. (\ref{cn}), (\ref{f}) and (\ref{del}) we obtain 
\beq
 {\overline c}_N (x) = f_D (x)
\label{cnf}
\eeq

Similar expressions can be found for the charm quark distribution in
the nucleon, 
$c_N (x)$. The above equation shows that 
the charm quark distributions in the nucleon are directly proportional to 
$g^2_{ND\Lambda_c}$. The value used  in ref. \cite{mt} was 
$\frac{g_{ND\Lambda_c}}{\sqrt{4\pi}}\,
=\,\frac{g_{\pi NN}}{\sqrt{4\pi}}\,=\, -3.795 $.   

Plugging the above calculated ${\overline c}_N\,(x)$, toghether with 
parton distributions of all flavors, in the standard DIS cross sections
expressions, $d\,\sigma\,/\,d\,x\,d\,Q^2$, Melnitchouk and Thomas were
able to reproduce the observed rise of the cross sections at large values  
of $M$ ($M \simeq 200\,GeV$) and $Q^2$.  

In this work we shall concentrate our attention on the coupling constant 
$g_{ND\Lambda_{c}}$, since it enters directly in the calculation of the
DIS cross section. The value used to reproduce the HERA events in \cite{mt}
, while a reasonable guess, has neither a solid theoretical justification
nor a phenomenological basis. It may, however, be calculated in the
framework of QCD Sum Rules (QCDSR). In the literature there are many 
calculations   of coupling constants with QCDSR. We shall make reference to
two of them: to \cite{rry}, which describes the general method and to 
\cite{cch}, 
which is recent and estimates $g_{NK\Lambda}$.

In order to calculate the $N$ $D^-$ $\Lambda_c$ coupling constant using
the QCD sum rule approach \cite{svz} we consider the three-point 
function
\begin{equation}
A(p,p^\prime,q)=\int d^4xd^4y \langle 0|T\{\eta _{\Lambda_c}(x)
j_5(y)\overline{\eta }_N(0)\}|0\rangle 
e^{ip^\prime x}e^{-iqy}\; , 
\label{cor}
\end{equation}
constructed with two baryon currents, $\eta _{\Lambda_c}$ and $\eta _N$,
for $\Lambda_c$ and the nucleon respectively, and the pseudoscalar meson
$D$ current, $j_5$, given by \cite{ioffe,dosch}
\beq
{\eta }_{\Lambda _c}= \varepsilon _{abc}({u}_a^TC\gamma _5{d}_b)Q_c \; ,
\eeq
\beq
{\eta }_N= \varepsilon _{abc}({u}_a^TC\gamma^\mu{u}_b)\gamma_5\gamma_\mu d_c 
\; ,
\eeq
\beq
j_5=\overline{Q} i\gamma_5u \; ,
\eeq
where $Q$, $u$ and $d$  are the charm, up and down quark fields
respectively, $C$ is the charge conjugation matrix.

Due to restrictions from Lorentz, parity and charge conjugation invariance
the general expression for $A(p,p^\prime,q)$ in Eq.(\ref{cor}) has the form
\beqa
A(p,p^\prime,q)&=&F_1(p^2,p'^2,q^2)i\gamma_5 +F_2(p^2,p'^2,q^2)\rlap{/}{q}
i\gamma_5
\nonumber\\*[7.2pt]
&+&F_3(p^2,p'^2,q^2)\rlap{/}{P}i\gamma_5+ F_4(p^2,p'^2,q^2)\sigma^{\mu\nu}
\gamma_5 p_\mu p^\prime_\nu \; ,
\label{struc}
\eeqa
where $q=p^\prime-p$ and $P=(p+p^\prime)/2$.

In the phenomenological side the different Lorentz structures appearing in 
Eq.(\ref{struc}) are obtained by the consideration of the $\Lambda_c$ and 
$N$ states contribution to the matrix element  in Eq.~(\ref{cor}):
\begin{equation}
\langle 0|\eta _{\Lambda_c}|\Lambda_c(p^\prime)\rangle\langle\Lambda_c
(p^\prime)|j_5|
N(p)\rangle\langle N(p)|\overline\eta_N|0\rangle \; ,
\label{j5}
\end{equation}
where the matrix element of the pseudoscalar current defines the pseudoscalar 
form-factor
\begin{equation}
\langle\Lambda_c(p^\prime)|j_5|N(p)\rangle=g_P(q^2)\overline{u}(p^\prime)i
\gamma_5
u(p)\; ,
\label{ps}
\end{equation}
where $u(p)$ is a Dirac spinor and $g_P(q^2)$ is related to $g_{ND\Lambda_c}$
through the relation \cite{rry}
\beq
g_P(q^2)={2m_D^2 f_D\over(m_u+m_c)}{g_{ND\Lambda_c}\over q^2-m_D^2} \; ,
\label{g}
\eeq
where $m_D$ and $f_D$ are the meson $D$ mass and decay constant and
$m_{u(c)}$ is the $u(c)$ quark mass. 
 
The other matrix elements contained in Eq.(\ref{j5}) are
of the form
\begin{eqnarray}
\langle 0|\eta _{\Lambda_c}|\Lambda(p^\prime)\rangle &=&\lambda_{\Lambda_c} 
u(p^\prime) 
\label{lamc}
\\
\langle N(p)|\overline\eta_N|0\rangle &=& \lambda_N
\overline{u}(p)\; ,
\label{lam}
\end{eqnarray}
where $\lambda_{\Lambda_c}$ and $\lambda_N$ are
the couplings of the currents with the respective hadronic states.

Saturating the correlation function Eq.(\ref{cor}) with
$\Lambda_c$ and $N$ intermediate states, and using Eqs.(\ref{j5}),
(\ref{ps}), (\ref{g}),  (\ref{lamc}) and (\ref{lam}) we get
\begin{eqnarray}
A^{(phen)}(p,p^\prime,q) &=&\lambda_{\Lambda_c}\lambda_N {2m_D^2 f_D
\over(m_u+m_c)}{g_{ND\Lambda_c}\over q^2-m_D^2}
{(\rlap{/}{p^\prime}+M_{\Lambda_c})\over p'^2-
M_{\Lambda_c}^2} i\gamma_5{(\rlap{/}{p}+M_N)\over p^2-M_N^2} + 
\mbox{higher resonances}\; ,
\label{aphen}
\end{eqnarray}
which can be rewritten as
\begin{eqnarray}
A^{(phen)}(p,p^\prime,q) &=&\lambda_{\Lambda_c}\lambda_N {2m_D^2 f_D
\over(m_u+m_c)}{g_{ND\Lambda_c}\over q^2-m_D^2}
{1\over p'^2-M_{\Lambda_c}^2}{1\over p^2-M_N^2}\left[(M_{\Lambda_c}M_N-p.
p^\prime)i\gamma_5\right.
\nonumber \\*[7.2pt]
&+&\left. {M_{\Lambda_c}+M_N\over2}\rlap{/}{q}i\gamma_5 + (M_{\Lambda_c}-M_N)
\rlap{/}{P}i\gamma_5-\sigma^{\mu\nu}\gamma_5p_\mu p^\prime_\nu \right] + 
\mbox{higher resonances}\; ,
\label{ficor}
\end{eqnarray}
where we clearly see all the Lorentz structures present in Eq.(\ref{struc}).
We will follow refs.\cite{rry,cch} and write a sum rule for the structure
$\rlap{/}{q}i\gamma_5$. As we are interested in the value of the
coupling constant at $q^2=0$, we will make a Borel transform to
both $p^2={p^\prime}^2\rightarrow M^2$. The contribution from excited
baryons will be taken into account as usual through the standard form
of ref.\cite{ioffe1}.

As in any QCD sum rule calculation, our goal is to make a match between the 
two representations
of the correlation function (\ref{cor}) at a certain region of
$M^2$: the OPE side and the phenomenological side. The contributions to the
OPE side are represented in figures 1, 2 and 3.

In the OPE side only odd dimension operators contribute to the
$\rlap{/}{q}i\gamma_5$ structure, since the dimension of Eq.(\ref{cor})
is four and $\rlap{/}{q}$ takes away one dimension. Therefore, the
perturbative diagram in Fig. 1 contributes through the $m_c$ operator.
To evaluate the perturbative contribution we write a double dispersion relation
to the amplitude $F_2$ in Eq.(\ref{struc}) and use the Cutkosky's rules
to evaluate the double discontinuity (see ref.\cite{ioffe2}). After 
Borel transforming with respect to $P^2=-p^2=-{p^\prime}^2$, and
subtracting the continuum contribution, we get
\beq
\left[\tilde{F_2}(M^2,q^2)\right]_1=-{1\over4\pi^2}\int_{m_c^2}^{u_0}du
\int_0^{u-m_c^2}ds\rho(u,s,q^2){1\over s-u}(e^{-u/M^2}-e^{-s/M^2})\; ,
\label{d1}
\eeq
with
\beqa
\rho(u,s,q^2)&=&{3\over2}{m_c\over(2\pi)^2}{1\over\sqrt{\lambda(s,u,q^2)}}
\int_0^{s}
dm^2\;\left\{ m^2\left({1\over4}+{\sqrt{s}p_0^\prime\over2\lambda(s,u,q^2)}
(u-p_0^\prime\sqrt{s}-m_c^2)\right)\right.
\nonumber \\*[7.2pt]
&-&\left.{m^4\over2}\left[{1\over2s}+{s\over\lambda(s,u,q^2)}\left(1-
{p_0^\prime\over\sqrt{s}}\right)^2\right]\right\}\Theta((\overline
{\cos\theta_K})^2-1)\; ,
\label{rho}
\eeqa
where $p_0^\prime=(s+u-q^2)/(2\sqrt{s})$ and
\beq
\overline{\cos\theta_K}=2s{u+m^2-m_c^2-p_0^\prime(s+m^2)/\sqrt{s}
\over(s-m^2)\sqrt{\lambda(s,u,q^2)}}\; .
\eeq
In Eq.(\ref{d1}) $\tilde{F_2}$ stands for the Borel transformation
of the amplitude $F_2$, the subscript 1 refers to the diagram 1 and
$u_0$ gives the continuum threshold for $\Lambda_c$.

The next
lowest dimension operator is the quark condensate with dimension three
(Fig. 2). In ref.\cite{rry} only the diagram shown in Fig. 2a was
considered because the pion mass was neglected and, therefore, only
terms proportional to $1/q^2$ contributed. Since in our case $m_D$ will
not be neglected we have also to consider the diagrams in Figs. 2b and 2c.
Their contributions are given by
\beq
\left[\tilde{F_2}(M^2,q^2)\right]_{2a}=-{\langle\overline{q}q\rangle\over8
\pi^2}{1\over q^2-m_c^2}M^4E_1\; ,
\label{d2a}
\eeq
with $E_1=1-e^{-u_o/M^2}(1+u_0/M^2)$.
\beq
\left[\tilde{F_2}(M^2,q^2)\right]_{2b+2c}={\langle\overline{q}q\rangle\over8
\pi^2}\int_{m_c^2}^{u_0}du\int_0^{u-m_c^2}ds{(2-\delta+\alpha)s+(1+\alpha)u
\over(u-s)^2}(e^{-u/M^2}-e^{-s/M^2})\; ,
\label{d2bc}
\eeq
where $\delta=1/2-\alpha(u+s)/(2s)$ and
\beq
\alpha={2s\over(u-s)^2}\left(m_c^2+{s-u\over2}\right)\; .
\eeq

The last class of diagrams which we will consider is the dimension 7 operators
of the type 
$m_c\,\langle\overline{q}q\overline{q}q\rangle$ ($q$ being a $u$
or $d$ quark) shown in Fig. 3. Other dimension 7 operators come from graphs
which contain at least one loop and are suppressed by factors $1/4\pi^2$.
The expressions for these contributions are
\beq
\left[\tilde{F_2}(M^2,q^2)\right]_{3a+3b}={\langle\overline{q}q\overline{q}q
\rangle\over6}{m_c\over q^2-m_c^2}\; ,
\label{d3ab}
\eeq
\beq
\left[\tilde{F_2}(M^2,q^2)\right]_{3c}=-{\langle\overline{q}q\overline{q}q
\rangle\over6}{1-e^{-m_c^2/M^2}\over m_c}\; .
\label{d3c}
\eeq

The Borel transformation of the phenomenological side gives
\beq
\left[\tilde{F_2}(M^2,q^2)\right]_{phen}=\lambda_{\Lambda_c}\lambda_N 
{m_D^2 f_D\over(m_u+m_c)}{g_{ND\Lambda_c}\over q^2-m_D^2}{M_{\Lambda_c}+
M_N\over M_{\Lambda_c}^2-M_N^2}(e^{-M_N^2/M^2}-e^{M_{\Lambda_c}^2/M^2})\; .
\label{fen}
\eeq
For $\lambda_{\Lambda_c}$ and $\lambda_N$ we use the values obtained from
the respective mass sum rules for the nucleon \cite{ioffe,rry} and for
$\Lambda_c$ \cite{dosch}:

\beqa
|\lambda_{\Lambda_c}|^2e^{-M_{\Lambda_c}^2/M^2}&=&{m_c^4\over512\pi^4}
\int_{m_c^2}^{u_o}e^{-u/M^2}\left[\left(1-{m_c^4\over u^2}\right)\left(1-
{8u\over
m_c^2}+{u^2\over m_c^4}\right)-12\ln\left({m_c^2\over u}\right)\right]
\nonumber \\*[7.2pt]
&+&{\langle\overline{q}q\overline{q}q\rangle\over6}e^{-m_c^2/M^2} \; ,
\label{rslc}
\eeqa
\beq
|\lambda_N|^2e^{-M_N^2/M^2}={M^6\over32\pi^4}E_2+{2\over3}\langle\overline{q}
q\overline{q}q\rangle\; ,
\label{rsn}
\eeq
where $E_2=1-e^{-s_0/M^2}(1+s_0/M^2+s_0^2/(2M^4))$ accounts for the continuum 
contribution with $s_0$ being the continuum threshold for the nucleon. We have
neglected the contribution of the gluon condensate in the mass sum 
rules and in the three point function since it is of little
influence.

In order to obtain $g_{ND\Lambda_c}$ we 
identify eq. (\ref{fen}) with the sum of eqs. 
(\ref{d1}, \ref{d2a}, \ref{d2bc}, \ref{d3ab}, \ref{d3c})
and using eqs. (\ref{rslc}, \ref{rsn}) we
solve the resulting equation for  $g_{ND\Lambda_c}$ as
a function of the Borel mass squared. The resulting curves are shown in 
Figs. 4 and 5.  Since eqs. (\ref{rslc}, \ref{rsn}) determine only the absolute 
value of $\lambda_N$ and $\lambda_{\Lambda_c}$ we can not determine the 
sign of $g_{ND\Lambda_c}$.

In this calculation there is no free parameter. There are, instead,
some numbers which are heavily constrained by other theoretical or 
phenomenological analyses. They are the sources of uncertainties in our
result. The quark condensate was taken to be  
$\langle\overline{q}q\rangle\,=\,-(0.23)^3\,\GeV^3$. The continuum  
thresholds  
appearing in $E_1$ and $E_2$ were chosen to be respectively 
$u_0\,=\,(M_{\Lambda_c}+0.7)^2\,\GeV^2$ and 
$s_0\,=\,(M_N+0.7)^2\,\GeV^2$. The hadron 
masses are $M_N\,=\,0.938\,\GeV$, $M_{\Lambda_c}\,=\,2.285\,\GeV$ and 
$m_D\,=\,1.8\,\GeV$. The charm quark mass was taken to be $m_c\,=\,1.3$ and
$1.5\,\GeV$. In order to take into account possible deviations from the
factorization hypothesis  \cite{narison}
we introduce a correction factor $K$ in the formula
\beq
\langle\overline{q}q\overline{q}q
\rangle\,=\,K
\langle\overline{q}q\rangle^2
\eeq
and consider the cases $K\,=\,1$ and  $K\,=\,2$. Finally we need
 the $D$ meson
decay constant $f_D$. This constant will be probably evaluated in the 
next generation of experiments on fully leptonic $D$ decays. So far it
has been usually estimated \cite{ratti} to be  $f_D=200\pm30$ MeV. On the
other hand, new measurements of $f_{D_s}$ \cite{cleo} by the CLEO 
Collaboration 
point to $f_{D_s}=344\pm37$ MeV. Considering that  $f_D$ and  $f_{D_s}$ are
related by $f_{D_s}/f_D\simeq1.15-1.2$ \cite{ratti} this indicates that 
$f_D\simeq300$ MeV. In view of this we take  $f_D$ in the interval 
$200 - 300$ MeV. 

The relevant Borel mass here is $M\simeq\frac{M_N+M_{\Lambda_c}}{2}$ and 
therefore we analyse the sum rule in the interval $1.5 \leq M^2 \leq 3.5$ 
GeV$^2$. In Fig. 4 we plot $g_{ND\Lambda_c}$ as a function  of $M^2$ for 
different values of $f_D$ and $K$. As it can be seen, the curve is rather 
flat showing that, for QCD sum rules standards, this sum rule result is 
stable. Moreover, the uncertainties of $50\%$ both in $f_D$ and   $K$ lead 
to a spread in the value of $g_{ND\Lambda_c}$. In Fig. 5 we
fix $f_D=250$ MeV and $K=1$ and vary $m_c$ from $1.3$ (solid line) to 
$1.5$ GeV (dashed line). The corresponding variation in   $g_{ND\Lambda_c}$ 
is of the order of $15\%$. We have also varied the continuum thresholds in
the interval $ M_N+0.5 \leq s_0 \leq   M_N+1.0 $ GeV and   
$ M_{\Lambda_c}+0.5 \leq u_0 \leq   M_{\Lambda_c}+1.0 $ GeV. Under these 
variations the  final result does not change appreciably. This indicates that 
the continuum contribution is under control. 

Considering all the uncertainties discussed above our final result is
\beq
{|g_{ND\Lambda_c}|\over\sqrt{4\pi}} \,=\,1.9\,\pm\,0.6
\eeq

This value is smaller than the one used in the calculations of ref. \cite{mt}
by a factor two. At first sight, since the normalization of the  $\overline c$ 
distribution in the nucleon is proportional to the square of  
$g_{ND\Lambda_c}$, using our value would significantly  reduce the 
contribution of the non-perturbative nucleon charm sea employed to study the
HERA events. However there are many other possible intermediate states apart
from $D-\Lambda_c$ which contribute to generate this charm sea like, for 
example, $D^*-\Lambda_c$, $D-\Sigma$ and $D^*-\Sigma$. These higher mass 
states are neglected in some calculations but, as shown in ref. 
\cite{pnndb}, already in the strange sector they may give important 
contributions for the evaluation of strange form factors and strange quark
momentum distributions. In the charm sector we expect the contribution
of these states to be even more important. Our conclusion is therefore, 
that the non-perturbatively generated charm sea remains qualitatively a good
candidate to explain the observed exotic events. Quantitative calculations
should, however, include the contribution of  other charm mesonic clouds,
specially in view of the small value found here for $g_{ND\Lambda_c}$. 
\vspace{1cm}

\underline{Acknowledgements}: This work has been supported by FAPESP
and CNPq.  We would like to warmly thank H.G. Dosch for 
instructive discussions. 

\vspace{1cm}
\newpage
\noindent
{\bf Figure Captions}\\
\begin{itemize}

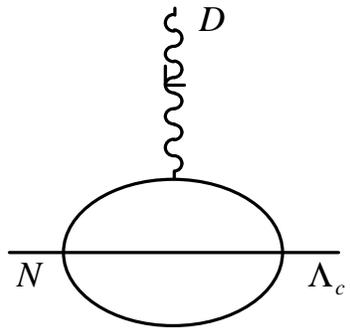
\item[{\bf Fig. 1}] Perturbative contribution to the OPE. 

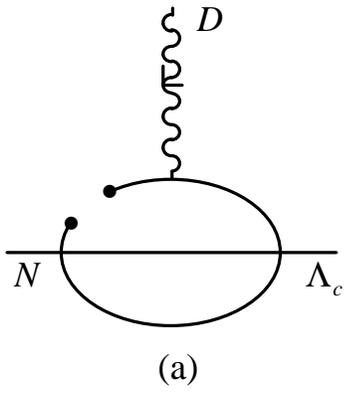
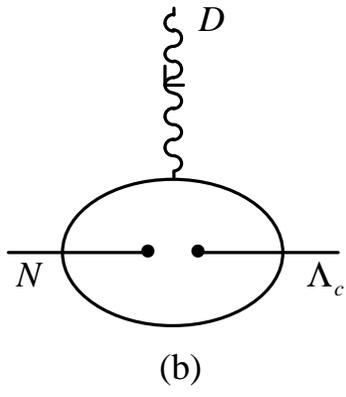
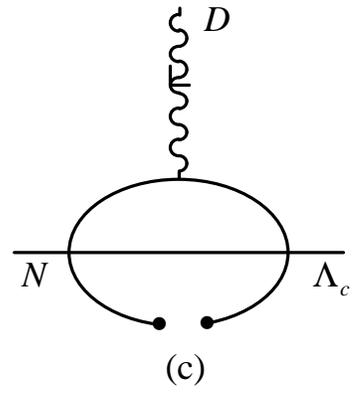
\item[{\bf Fig. 2}] Diagrams that contribute to the Wilson coefficient of the
operator $ {\overline q} q$. 

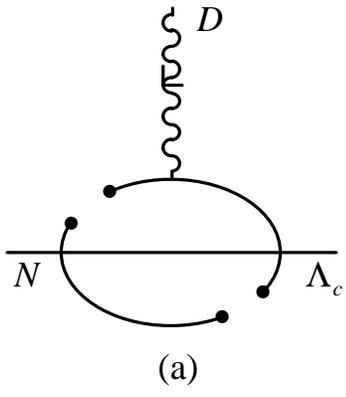
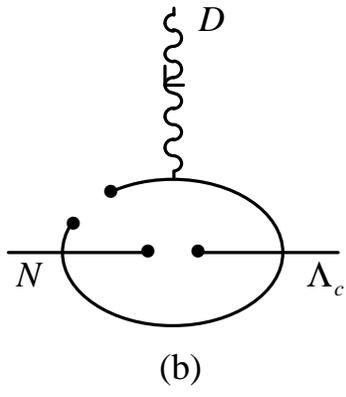
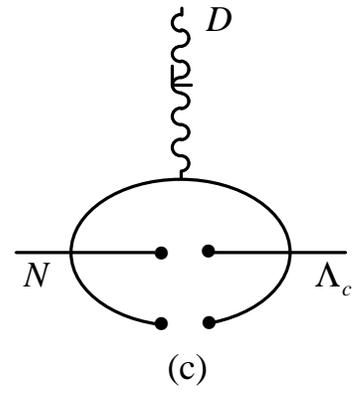
\item[{\bf Fig. 3}] Diagrams that contribute to the Wilson coefficient of the 
operator $ {\overline q} q   {\overline q} q$. 

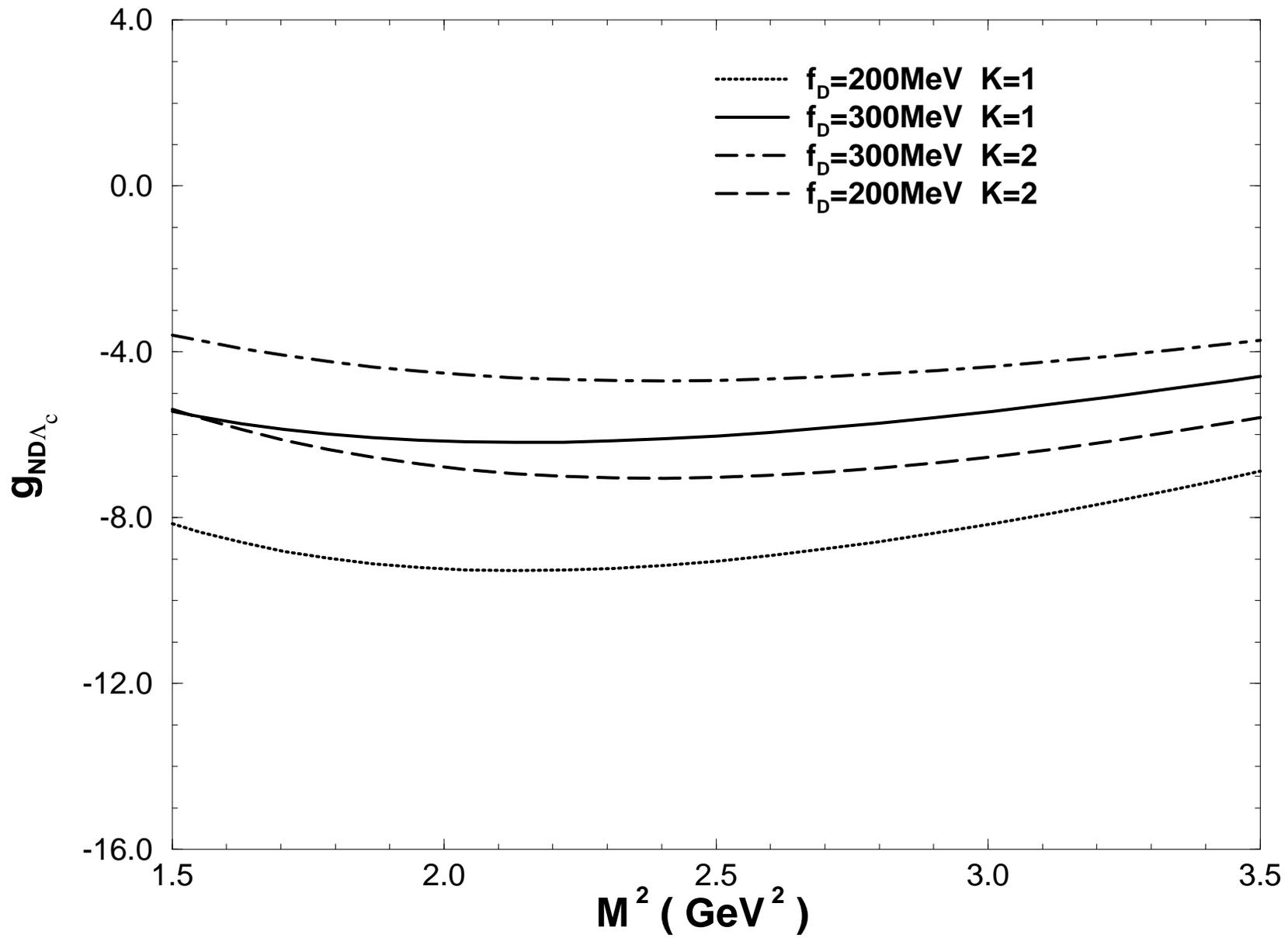
\item[{\bf Fig. 4}] $g_{ND\Lambda_c}$ as a function of the squared Borel mass
$M^2$. The different lines correspond to different choices for the $D$ meson
decay constant, $f_D$,  and for the factorization correction factor $K$ as
indicated in the legends.

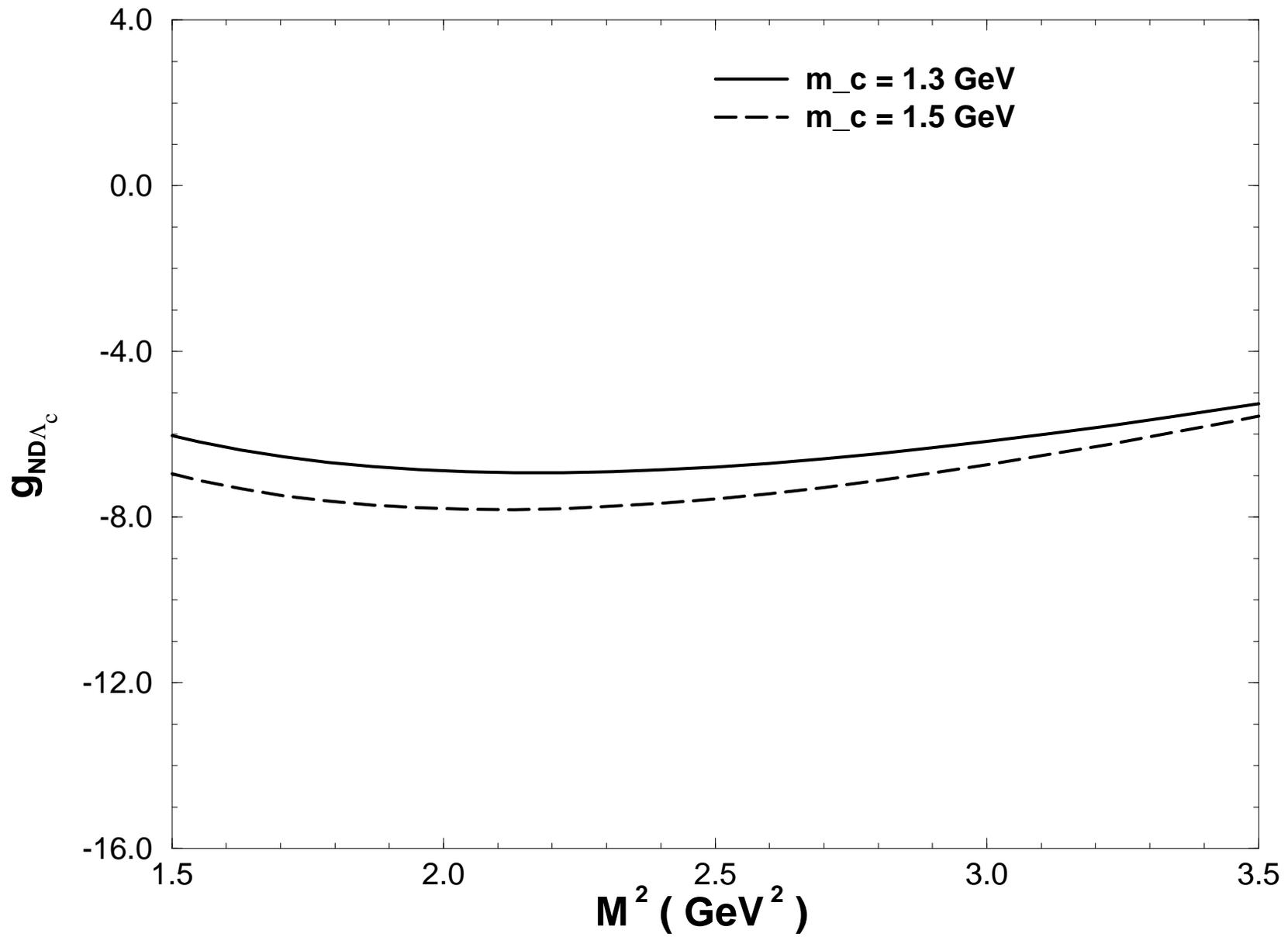
\item[{\bf Fig. 5}] The same as Fig. 4 for diferent values of the $c$ quark
mass. $f_D= 250$ MeV and $K=1$.

\end{itemize}

\end{document}